\documentclass[10pt,english,journal]{IEEEtran}
\usepackage[T1]{fontenc}
\usepackage[latin9]{inputenc}
\usepackage{geometry}
\usepackage{amsmath}
\usepackage{xpatch}
\usepackage{amssymb}
\usepackage{esint}
\usepackage{mathtools}
\usepackage{amsthm}
\usepackage{nicefrac}
\usepackage{bigints}
\usepackage{mathtools}
\usepackage{blkarray, bigstrut}
\usepackage{physics}
\usepackage{calligra}
\usepackage{graphicx}
\usepackage{epstopdf}
\usepackage{dsfont}
\usepackage{array,ragged2e}
\usepackage{enumitem}
\usepackage{etoolbox}
\usepackage{babel}
\usepackage[nopar]{lipsum}
\usepackage{psfrag}
\usepackage[ruled,lined,linesnumbered]{algorithm2e}
\usepackage{algorithmic}
\usepackage{algorithm2e}
\usepackage{balance}
\usepackage{float}
\usepackage{subcaption}

\SetKw{KwBy}{by}

\makeatletter
\newcommand{\removelatexerror}{\let\@latex@error\@gobble}
\makeatother

\graphicspath{ {Figures/} }

\xpatchcmd{\proof}{\hskip\labelsep}{\hskip5\labelsep}{}{}  %% change 5 here as you wish
\makeatletter
\xpatchcmd{\proof}{\@addpunct{.}}{\@addpunct{:}}{}{}
\makeatother

\renewcommand\[{\begin{equation}}
\renewcommand\]{\end{equation}} 
\usepackage{listings}
\usepackage{fancyvrb}
\usepackage{framed}

\usepackage{courier}
\usepackage[usenames,dvipsnames,table]{xcolor}

\definecolor{dkgreen}{rgb}{0,0.3,0}
\definecolor{gray}{rgb}{0.5,0.5,0.5}

\makeatletter
\newcommand*{\rom}[1]{\expandafter\@slowromancap\romannumeral #1@}
\makeatother

\usepackage{siunitx}
\usepackage{tabu}
\usepackage{booktabs}
\usepackage{multirow}
\usepackage{capt-of}
\usepackage{array}
\usepackage{arydshln}
\newcommand{\comment}[1]{}
\begin{document}
\title{Towards Quantum-Enabled 6G Slicing}
\author{Farhad Rezazadeh$^{1,2}$, Sarang Kahvazadeh$^1$, Mohammadreza Mosahebfard$^{2,3}$\\
{\normalsize{} $^1$ Telecommunications Technological Center of Catalonia (CTTC), Barcelona, Spain\\ $^2$ Technical University of Catalonia (UPC), Barcelona, Spain}\\  $^3$ i2CAT Foundation, Barcelona, Spain\\
{\normalsize{}Contact Emails: \texttt{\{frezazadeh, skahvazadeh\}@cttc.es, reza.mosahebfard@i2cat.net}}\vspace{-5mm}}
\maketitle
\thispagestyle{empty}

\section{Extended Abstract}
\begin{figure}[!t]
\centering
\includegraphics[clip, trim = 0cm 0cm 0cm 0cm, width=0.4\textwidth]{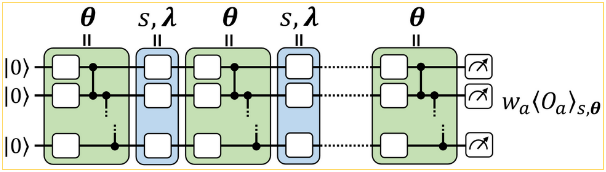}
\caption{\small The generic layout of PQC with data re-uploading. State $s$, action $a$, observable weights w, input scaling parameters $\lambda$, and function approximator $\langle O_a \rangle_{s, \theta}$ \cite{skolik}.}
\label{fig:VQC}
\vspace{-7mm}
\end{figure}

The quantum machine learning (QML) paradigms and their synergies with network slicing can be envisioned to be a disruptive technology on the cusp of entering to era of sixth-generation (6G), where the mobile communication systems are underpinned in the form of advanced tenancy-based digital use-cases to meet different service requirements \cite{masssive-slicing,chapter_far,moh-reza,moh-reza2,mano-far}. To overcome the challenges of massive slices such as handling the increased dynamism, heterogeneity, amount of data, extended training time, and variety of security levels for slice instances, the power of quantum computing pursuing a distributed computation and learning can be deemed as a promising prerequisite. In this intent, we propose a cloud-native federated learning framework based on quantum deep reinforcement learning (QDRL) where distributed decision agents deployed as micro-services at the edge and cloud through Kubernetes infrastructure then are connected dynamically to the radio access network (RAN). Specifically, the decision agents leverage the remold of classical deep reinforcement learning (DRL) algorithm into variational/parametrized quantum circuits (VQCs or PQCs) to obtain the optimal cooperative control on slice resources. 
% Network slicing is the embodiment of severing the network into different segments that enables the multiplexing of virtualized and isolated logical networks---or slices---on top of the same physical network infrastructure. This paradigm is a paramount feature in beyond 5G (B5G)/6G systems that leverages network softwarization and virtualization technologies such as software-defined networking (SDN) and network function virtualization (NFV) \cite{moh-reza,moh-reza2}. Indeed, it provides the necessary programmability and flexibility to manage the life cycle of chained virtual network functions (VNFs) \cite{chapter_far}. It is envisaged that 6G will involve massive slices, which are made up of numerous micro or macro services \cite{masssive-slicing}.
%Given the cloud nature of these resources, the networking resources associated to each slice can be dynamically orchestrated and tailored to meet the performance requirements of running services.
In this context, temporal variations of the traffic demand deeply complicate resource planning and allocation tasks, especially in the RAN domain. Hence, the complexity of automated management and orchestration (MANO) \cite{mano-far} operations such as resource allocation arises dramatically concerning these progressive developments. 
Indeed, developing new DRL approaches to tackle these challenging multi-tasks can be considered a necessity in multi-domain 6G. A trend is emerging to combine DRL with quantum computing in the form of QDRL that has significantly fewer parameters by leveraging the power of quantum neural network (QNN) and utilizing the qubit properties such as superposition and entanglement. Considering the privacy concerns associated with large-scale and distributed 6G infrastructures and the limited quantum capabilities of different quantum machines in the noisy intermediate-scale quantum (NISQ) \cite{Funcke} era, we propose a federated learning scheme to provide a distributed computing pursuing a secure QML training while enriching the capabilities of the QDRL decision agents through learning more trajectory of experiences and model generalization. The suitability of utilizing VQCs for function approximation in RL is still an open question \cite{skolik}. As shown in Fig.~\ref{fig:VQC}, we implement decision agents based on a VQC in TensorFlow quantum (TFQ) \cite{Tensorflow_quantum}, where the Q-function approximator trained with deep Q-network (DQN). To obtain better expressive power \cite{Schuld}, we leverage re-uploadings \cite{Perez} for single-qubit encodings.

\begin{figure}[!h]
\vspace{-5mm}
\centering
\includegraphics[clip, trim = 0cm 0cm 0cm 0cm,width=4.5cm]{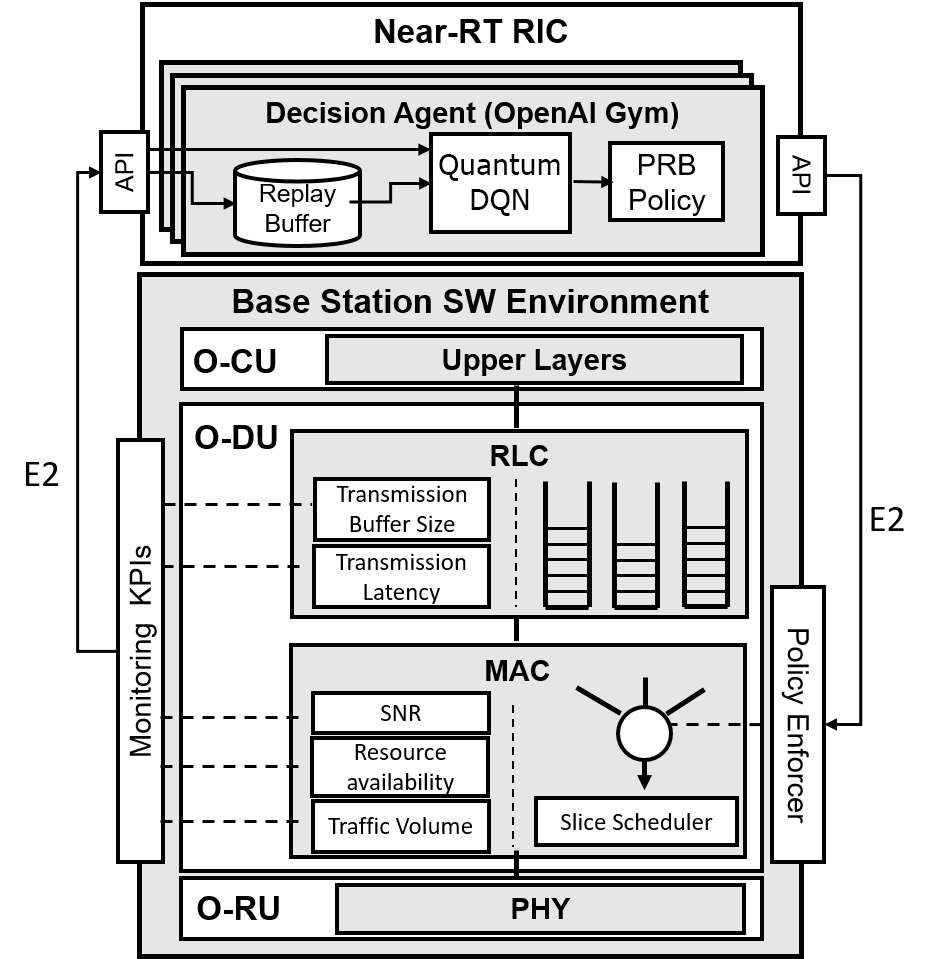}
\caption{\small The software protocol stack overview with O-RAN compliance for gNodeB.}
\label{fig:sw_architecture}
\vspace{-.3cm}
\end{figure}

\begin{figure}[!h]
\centering
\includegraphics[clip, trim = 0cm 0cm 0cm 0cm, width=0.4\textwidth]{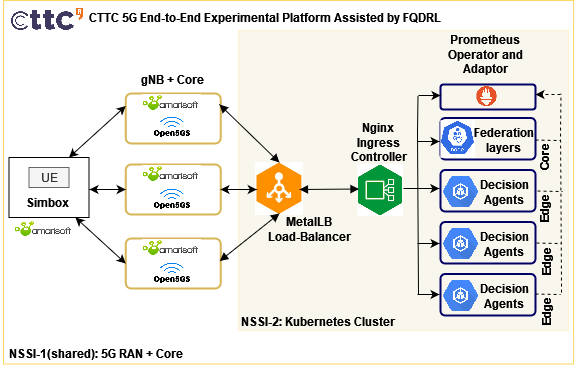}
\caption{\small The proposed testbed framework with QDRL-based decision agents.}
\label{fig:FQDRL}
\vspace{-3mm}
\end{figure}

In our proposed framework, quantum DQN (QDQN) \cite{qdrl_Le,qdrl_Yen,qdrl_Wei} agents optimally allocate radio resources to each slice, while a federation layer enables a periodical exchange of the QDQN's parameter values to improve the learning process across multiple agents of the same slice. We analyze the performance of proposed framework based on both simulation and testbed environment. As depicted in Fig.~\ref{fig:sw_architecture}, we implement our framework based on Python programming and exploiting OpenAI Gym library~\cite{Globe_far} interfacing QDRL agents with a custom base station (BS) simulator environment. We highlight the O-RAN \cite{ORAN_architecture} synergies with respect to our proposed approach which includes virtual transmission queues and main PHY/MAC/RLC functionalities, together with O-RAN  E2 interface to allow gathering the slice networking statistics from each distributed unit (O-DU).
The QDRL agents enforce physical resource blocks (PRBs) policy decisions in the BS slice scheduler and finally a federation layer connects the QDRL agents of the $i$-th slice to enable inter-agent information exchange and expedite the overall learning procedure.
Fig.~\ref{fig:FQDRL} demonstrates our testbed structure where three Amarisoft\footnote{https://www.amarisoft.com} gNodeBs jointly with a single Open5Gs\footnote{https://open5gs.org} core provide 5G segments and connectivity. In this 5G segment, Open5Gs core can be containerized in Kubernetes infrastructure or can be deployed in virtual machines (VMs). We also have Amarisoft user emulators (UEs) that can generate 5G users. As shown in Fig.~\ref{fig:FQDRL}, our 5G segment as a shared network slice sub-instance (NSSI-1) connected with our designed platform as a service (PaaS) Kubernetes infrastructure (according to ETSI-029 \cite{ETSI-029}) with scalable network sub instances (NSSI-2). PaaS Kubernetes infrastructure can be scaled in/out NSSIs for the sake of resource management. The decision agents are deployed in edge nodes while federation layers at the cloud node in our Kubernetes infrastructure connected to 5G segments. 
\vspace{-3mm}

\begin{figure}[!h]
\centering
\includegraphics[clip, trim = 0cm 0cm 0cm 0cm,width=6cm]{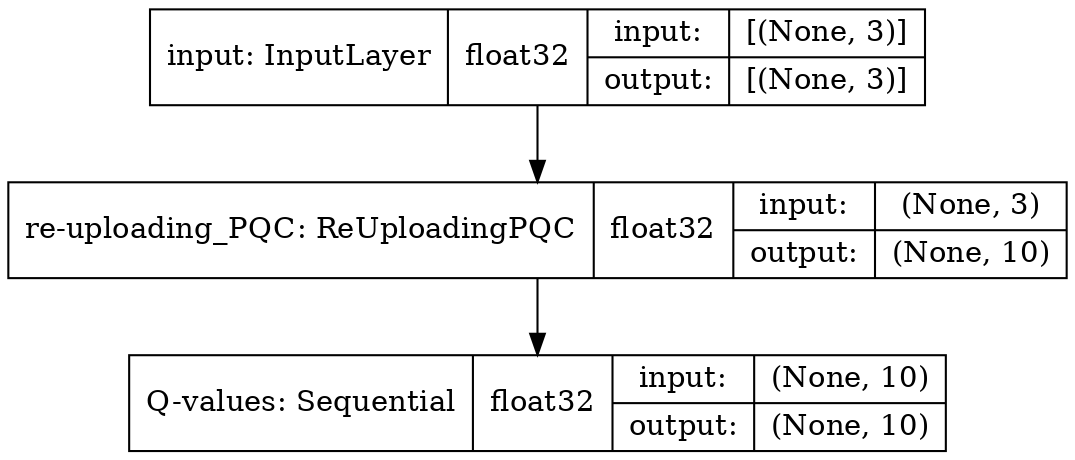}
\caption{\small The architecture of local agent model with 3 qubits.}
\label{fig:single_agent_architecture}
\vspace{-2mm}
\end{figure}

\begin{figure}[h]
\centering
\includegraphics[clip, trim = 0cm 0cm 0cm 0cm,width=5cm]{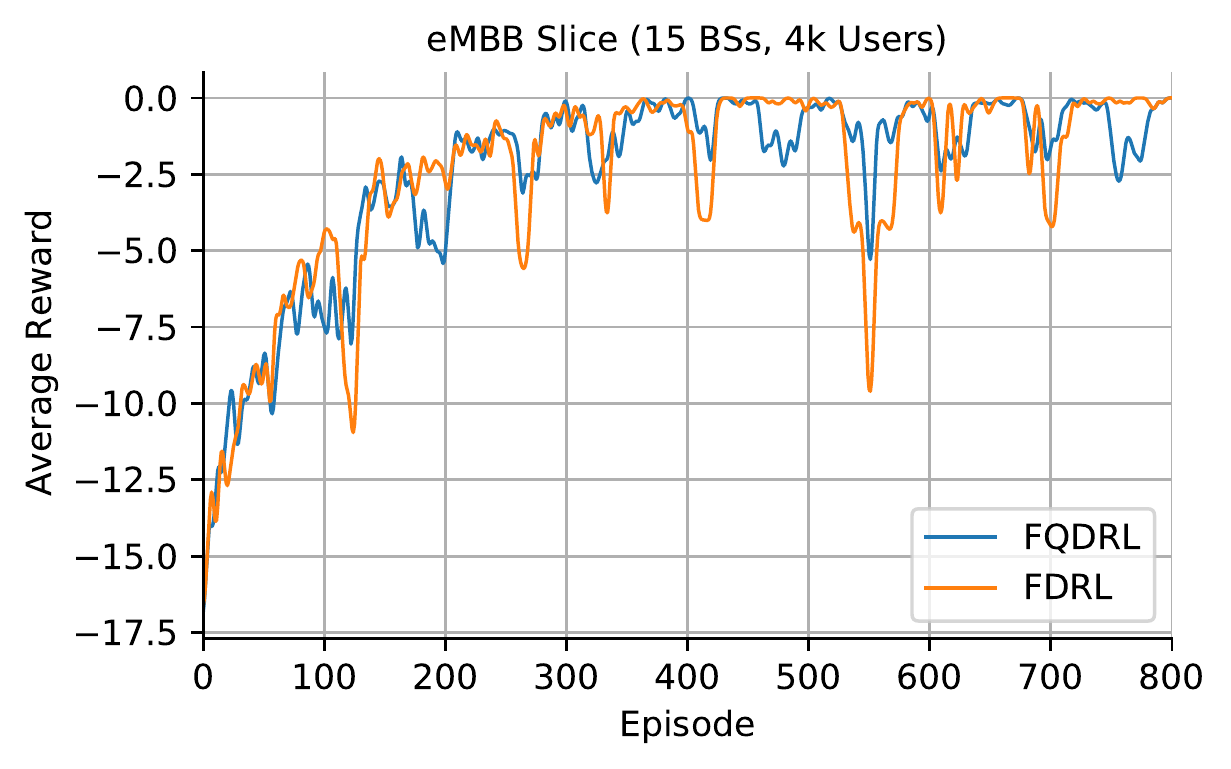}
\caption{\small Comparison of global performances for different federation approaches.}
\label{fig:reward_bench}
\vspace{-5mm}
\end{figure}

We perform simulations to compare the initial results of federated classical DRL (FDRL) and FQDRL to demonstrate the potential of FQDRL based on a standard RL benchmark from OpenAI Gym \cite{Globe_far}. We use a continuous state space while the action space is discrete and run our experiments on a dedicated server, equipped with two Intel(R) Xeon(R) Gold 5218 CPUs @ 2.30GHz and two NVIDIA GeForce RTX 2080 Ti GPUs. Moreover, the deep neural networks (DNNs) are implemented based on TensorFlow-gpu and utilize TensorFlow Quantum \cite{Tensorflow_quantum} and Cirq\footnote{https://quantumai.google/cirq}. Fig.~\ref{fig:single_agent_architecture} shows the architecture of a single FQDRL agent with 5 layers in the PQC. 
Fig.~\ref{fig:reward_bench} provides a comparison of learning performance in terms of average reward over 15 BS (gNBs).
\vspace{-2mm}
\section*{Acknowledgement}This work was partially funded by the Spanish Government (MICCIN) under Grant PCI2020-112049 and by the Electronic Components and Systems for European Leadership Joint Undertaking (JU) under grant agreement No 876868. This JU receives support from the EU`s H2020 research and innovation programme and Germany, Slovakia, Netherlands, Spain, Italy and also was supported by H2020 5GMediaHUB (Grant Agreement no.101016714) and TSI-063000-2021- 1- 6GSatNet-GS project.

\end{document}